\documentclass[3p,12pt,a4paper,english]{article}

\input{glyphtounicode}
\usepackage[sc]{mathpazo} 
\usepackage[expansion=true,protrusion=true]{microtype} 

\usepackage{tabularx}
\usepackage{verbatim}
\usepackage{natbib}
\usepackage{mathrsfs}
\usepackage{float}
\usepackage{amsfonts}
\usepackage{amssymb}
\usepackage{amsmath}
\usepackage{lscape}
\usepackage[pdftex]{color,graphicx}
\usepackage{rotating}
\usepackage{setspace}
\usepackage[margin=1in]{geometry}
\usepackage{url}
\usepackage{MnSymbol}
\usepackage{eurosym}
\usepackage{bbm}
\usepackage{subfig}
\usepackage{verbatim}
\usepackage{setspace}
\usepackage{lipsum}
\usepackage{longtable,threeparttable,pdflscape}
\usepackage{footmisc}
\usepackage{color,hyperref}
\hypersetup{pdfstartview={XYZ null
null 1.10},colorlinks=true, citecolor=blue, linkcolor =blue,
urlcolor=blue}
\usepackage[english]{babel}
\usepackage[utf8]{inputenc}
\usepackage{authblk}
\usepackage[T1]{fontenc}
\usepackage{babel}
\usepackage{tikz}
\usepackage{afterpage}

\usetikzlibrary{positioning,shapes.geometric}
\usepackage{multicol}
\newcolumntype{A}{>{\arraybackslash}m{4.5cm}}
\newcolumntype{B}{>{\arraybackslash}m{12cm}}
\newcolumntype{C}{>{\centering\arraybackslash}m{2.5cm}}
\newcolumntype{D}{>{\centering\arraybackslash}m{4cm}}
\newcolumntype{E}{>{\centering\arraybackslash}m{1cm}}
\newcolumntype{F}{>{\centering\arraybackslash}m{2cm}}
\newcolumntype{G}{>{\arraybackslash}m{2.5cm}}
\usepackage{authblk}
\usepackage{csquotes}
\MakeOuterQuote{"}

\begin{document}

\title{\vspace{-2cm}\textbf{Environmental-Social-Governance Preferences and Investments in Crypto-Assets}\footnote{We would like to thank to an anonymous referee, Martin Šuster, Ján Klacso and Roman Vasiľ for their helpful comments on an earlier version of this paper. The authors are solely responsible for the content of the paper. The views expressed are purely those of the authors and may not in any circumstances be regarded as stating an official position of the European Commission, National Bank of Slovakia, and the Oesterreichische Nationalbank. Any remaining errors are solely ours.}}

\author{Pavel Ciaian\footnote{Corresponding author. Joint Research Centre (JRC), European Commission, e-mail: \texttt{pavel.ciaian@ec.europa.eu}.} \hspace{0.5cm} Andrej Cupak\footnote{Research Department, National Bank of Slovakia; University of Economics in Bratislava, e-mail: \texttt{andrej.cupak@nbs.sk}.} \hspace{0.5cm} Pirmin Fessler\footnote{Economic Microdata Lab, Oesterreichische Nationalbank, e-mail: \texttt{pirmin.fessler@oenb.at}.} \hspace{0.5cm} d'Artis Kancs\footnote{Joint Research Centre (JRC), European Commission, e-mail: \texttt{d'artis.kancs@ec.europa.eu}.}}

\date{\normalsize{\today}}

\maketitle

\vspace{-0.5cm}\begin{abstract} Individuals invest in Environmental-Social-Governance (ESG)-assets not only because of (higher) expected returns but also driven by ethical and social considerations. Less is known about ESG-conscious investor subjective beliefs about crypto-assets and how do these compare to traditional assets. Controversies surrounding the ESG footprint of certain crypto-asset classes - mainly on grounds of their energy-intensive crypto mining - offer a potentially informative object of inquiry. Leveraging a unique representative household finance survey for the Austrian population, we examine whether investors' ESG preferences can explain cross-sectional differences in individual portfolio exposure to crypto-assets. We find a strong association between investors' ESG preferences and the crypto-investment exposure. The ESG-conscious investor attention is higher for crypto-assets compared to traditional asset classes such as bonds and shares.
\end{abstract}

\clearpage\newpage

\section{Introduction}
\label{sec:section1}

In a standard asset pricing framework, financial decisions are determined by investor's preferences and beliefs over asset returns. A more recent literature has identified also the relevance of investor environment and non-pecuniary effects in driving cross-sectional differences in investment decision \citep{chen2020,jiang2021}. Accordingly, an investor weighs between optimising a standard mean-variance utility and maintaining a ``target portfolio''. The mean-variance utility captures the pecuniary effect of standard mean-variance preferences; investors' characteristics and personality differences affect investment decisions through these channels of beliefs and risk preferences. The target portfolio, in a reduced form, reflects non-pecuniary effects, such as the social and ethical/moral concerns.

The focus of the present paper is on non-pecuniary effects related to Environmental-Social-Governance (ESG)\footnote{Through the paper we use term `ESG' in line with the extant literature despite we can observe and measure only the environmental (E) and social (S) attitudes of individuals.} preferences in retail investor portfolio exposure to crypto-assets. Indeed, little is known about ESG-conscious investor subjective beliefs about crypto-assets and how do these compare to traditional assets in the portfolio formation. We aim to answer the question to what extent can environmental and ethical considerations explain cross-sectional differences in crypto-asset investments after controlling for investor individual characteristics and demographic variables. To benchmark our results, we compare how investors' ESG preferences relate to portfolio exposure to crypto-assets on the one side and `ESG-blind' traditional financial assets, such as bonds and shares,\footnote{The survey questions do not identify separately ESG stocks and non-ESG assets.} on the other side.

This is the first paper that investigates if and to what extent ESG preferences drive individual portfolio exposure to crypto-assets by leveraging representative household-level portfolio data. The Austrian Survey of Financial Literacy (ASFL) data are unique in the sense that most of standard household finance surveys do not include crypto-asset holdings as separate items. The ASFL data allow us to distinguish between individuals' investment choices between crypto-assets, bonds and shares. A common empirical challenge when estimating the effect of ESG preferences on portfolio composition is the potential endogeneity of ESG preferences. We take a number of steps in response to endogeneity concerns including an alternative IV identification strategy proposed by \cite{lewbel2012}.

There are two strands of literature our work is related to. First, the household finance and asset pricing contributions in the sustainable and responsible investing (SRI) literature have examined the unconditional and conditional ESG stock return performance. The empirical literature has established that ESG assets outperform non-ESG assets when positive shocks hit the ESG factor, which captures for example shifts in consumers' tastes for green products and investors' tastes for green holdings \citep[e.g.][]{pastor2021a}. The explosive growth in responsible investing has given rise to a growing theoretical asset pricing literature that relies on non-pecuniary utility functions \citep[e.g.][]{pastor2021b,liu2022}. The conceptual explanation for the incorporation of ESG preferences into investment decision-making relies on the idea that social preferences can affect investment decisions because they serve as a proxy for value-relevant information or risk, they enhance performance or reduce risk \citep{krueger2020}. Empirically the link between ESG preferences and portfolio choice is not that clear. \cite{anderson2021} find no relationship between ESG attitudes and pro-environmental portfolios. Even less is known about non-pecuniary utility and its relation to crypto-assets. How do ESG-conscious investors value crypto-assets, and do sustainable crypto investment products offer superior risk-adjusted returns? Our study contributes to a better understanding of non-pecuniary effects in individual investment decisions by assessing the role of an ESG-driven motivation in individual crypto investment decisions and benchmarking results against traditional asset holdings.

Second, a rich crypto-asset literature estimates the realised ESG footprint of crypto-assets \citep[e.g.][]{barone2019,kohler2019,richman2021,teichmann2021,parmentola2022} or how pecuniary effects explain individuals' investment demand for crypto-assets \citep{bouri2019,xi2020}. On the one hand, this literature suggests that crypto-assets have the potential to generate a variety of social and governance benefits either directly via a decentralised governance mechanism or via the way crypto-assets and the underlying blockchain technology are deployed \citep[e.g.][]{ciaian2016,chapron2017,richman2021}. On the other hand, crypto-assets are sometimes associated with undesirable social activities, such as illicit trade, money laundering and tax evasion \citep[e.g.][]{barone2019,teichmann2021}. Further, due to a continuously growing energy consumption to maintain the underlying blockchain network, certain crypto-assets are associated with negative environmental impacts. Particularly the Proof-of-Work (PoW) consensus mechanism of Bitcoin consumes large amounts of energy generating negative environmental externalities \citep[e.g.][]{dilek2019,kohler2019}.

Overall, the literature findings of the relationship between social, environmental and governance aspects of crypto-assets on individual portfolio exposure to crypto holdings is largely inconclusive; it depends among others on the specific crypto-asset and individual perceptions of investors. Our main finding that stronger ESG preferences go along with higher probability to hold crypto-assets might seem somewhat surprising at first sight, however, it conceivably ties in with previous literature on ESG attitudes and financial portfolio choice finding that socially ``desirable'' preferences communicated do not always match the preferences revealed from portfolio choice \citep[see][]{anderson2021}.

The present study contributes to enhancing our knowledge about this interplay between stated preferences, revealed ESG beliefs and portfolio holdings by providing novel insights about the relationship between environmental \& social preferences and individual portfolio exposure to crypto-assets. Indirectly it therefore also conveys information about the perceived ESG footprint of crypto-assets by retail investors. Furthermore, it illustrates the value added of augmenting the information on crypto-assets in standard household finance surveys for enhancing our understanding about crypto-asset holdings and investment decisions within a general portfolio choice context and along with socio-economic information.

\section{Data and variables}
\label{sec:section2}

\subsection{Austrian Survey of Financial Literacy}

We leverage a unique household portfolio data from the Austrian Survey of Financial Literacy (ASFL) for 2019 - the Austrian contribution to the OECD/INFE survey on adult financial literacy. The standard OECD survey comprises questions on financial knowledge, attitudes and behaviour, used by the OECD to calculate the respective financial literacy scores, as well as several control variables and demographics \citep[see][]{oecd2018}. The ASFL survey was conducted with 1,418 respondents through computer-assisted personal interviews (CAPIs) between April and May 2019. After verifying individual responses and cleaning the data, the final working sample consists of 1,016 individual-level observations. The main descriptive results of the ASFL as well as methodological details are reported in \cite{fessler2020}. First results on crypto-assets owners in Austria are reported in \cite{stix2021}.

The description of variables used in empirical estimations is provided in Table \ref{table:tablea1} of the Appendix. Our main dependent variable measures whether an individual owns crypto-assets (\textit{Crypto-assets ownership}). To compare how investors' behaviour differs between crypto-assets and traditional financial assets, we construct two further dependent variables capturing individual' ownership of bonds (\textit{Bonds ownership}) and shares (\textit{Stocks/shares ownership}).

\begin{figure}[!htbp]
\caption{Distribution of ESG scores}
    \begin{center}
    \label{figure:fig1}
\footnotesize

\subfloat[ESG1 score]{\includegraphics[width=11cm,keepaspectratio]{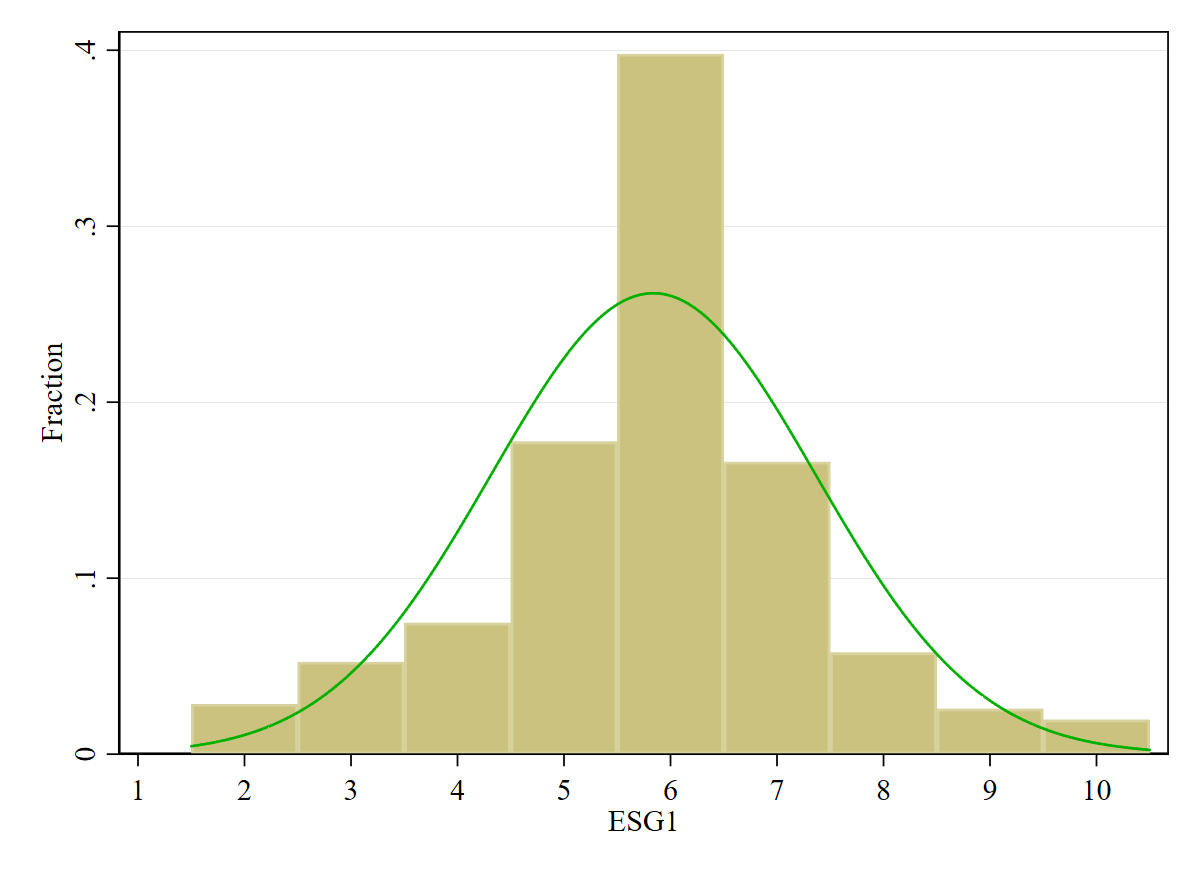}} \\
\subfloat[ESG2 score]{\includegraphics[width=11cm,keepaspectratio]{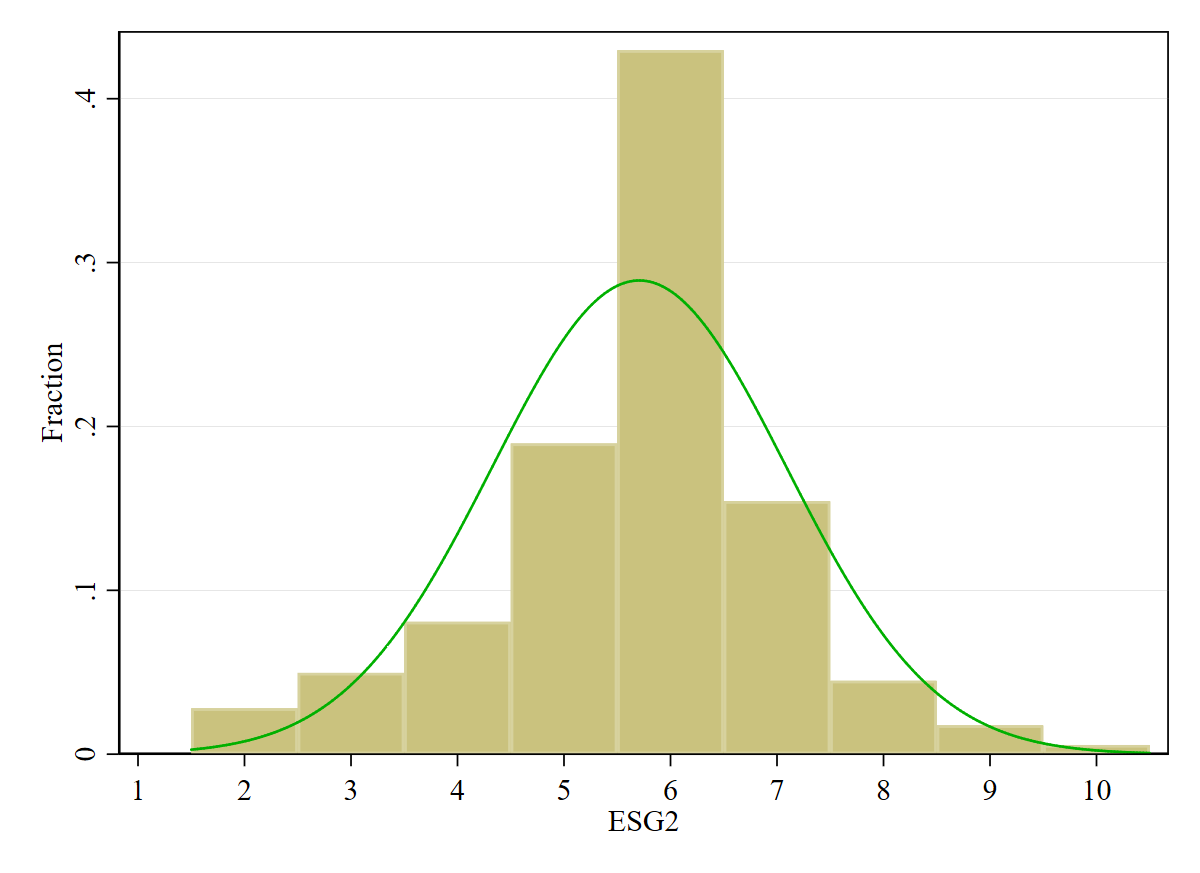}}

    \end{center}
\footnotesize{Note: This graph shows the distribution of two ESG scores overlaid by the normal density curve (green solid line).}
\newline \footnotesize{Source: ASFL 2019}
\end{figure}

The explanatory variables of particular interest are those capturing ESG preferences of retail investors. We consider one variable proxying environmental attitudes, \textit{Preferences for enviro. issues (E)}, and two alternative variables capturing social attitudes, \textit{Preferences for social issues (S1)} and \textit{Preferences for social issues (S2)}, respectively. All three ESG preference variables take scores between 1 to 5 with a higher value indicating stronger ESG attitude. We construct composite ESG indicators that measure combined environmental and social attitudes of surveyed individuals. The composite ESG indicators are constructed by summing up the values of environmental and social attitude variables: i.e. \textit{ESG1} is calculated as the sum of \textit{E} and \textit{S1} and \textit{ESG2} as the sum of \textit{E} and \textit{S2}. Distributions of the computed ESG scores are shown in Figure \ref{figure:fig1}.

\vspace{0.25cm}
\begin{center}
[Figure \ref{figure:fig1} around here]
\end{center}
\vspace{0.25cm}

Following previous studies on individual's portfolio composition and returns and risky financial behaviour \citep[e.g.][]{duarte2021,ehrlich2022}, we include a number of control variables to account for individual characterises such as age, gender, education (\textit{Primary education}, \textit{Secondary education}, \textit{Tertiary education}) and income (\textit{Individual monthly income}). An important driver of investment decisions of individuals identified in the literature is their objective financial literacy as well as their self-assessment of their own financial knowledge \citep[see][]{lusardi2014,bannier2018,bannier2019}. Two alternative explanatory variables describe financial literacy: the objectively measured financial literacy (\textit{Objective fin. literacy}) and the self-reported financial literacy (\textit{Confidence in own fin. knowledge}). In an attempt to control for risk attitudes of surveyed responders, which were identified in the literature to affect investment decisions \citep{jiang2021}, we also include a variable capturing self-reported willingness to take investment risk (\textit{Risk attitude score}).

\subsection{Descriptive statistics}

Table \ref{table:table1} reports descriptive statistics of surveyed individuals. Overall, around 3\% of surveyed individuals report holding crypto-assets, while the share of individuals owning bonds or shares is 7\% and 11\%, respectively\footnote{Note, that while there is some overlap between bond- and share-holders, it is far from perfect. About 62\% of those holding bonds hold also shares and about 40\% of those holding shares hold also bonds.}. The average score for environmental preferences (3.7) exceeds the social preferences scores (2.2 and 2.0, respectively) suggesting that the Austrian population might find environmental issues related to finance more important than social ones. Both the objective and subjective financial literacy scores (average values of 5.3 and 3.3) place Austria to a group of OECD countries with a high financial awareness \citep[see][for international comparison]{oecd2018}. Summary statistics of other relevant variables used in the empirical analyses are detailed in Table \ref{table:table1}.

\vspace{0.25cm}
\begin{center}
[Table \ref{table:table1} around here]
\end{center}
\vspace{0.25cm}

\begin{figure}[!h]
\caption{Correlation between environmental/social attitudes and holdings of different assets}
    \begin{center}
    \label{figure:fig2}
\footnotesize

\subfloat[Pr. of holding crypto-assets and ESG1 score]{\includegraphics[width=8cm,keepaspectratio]{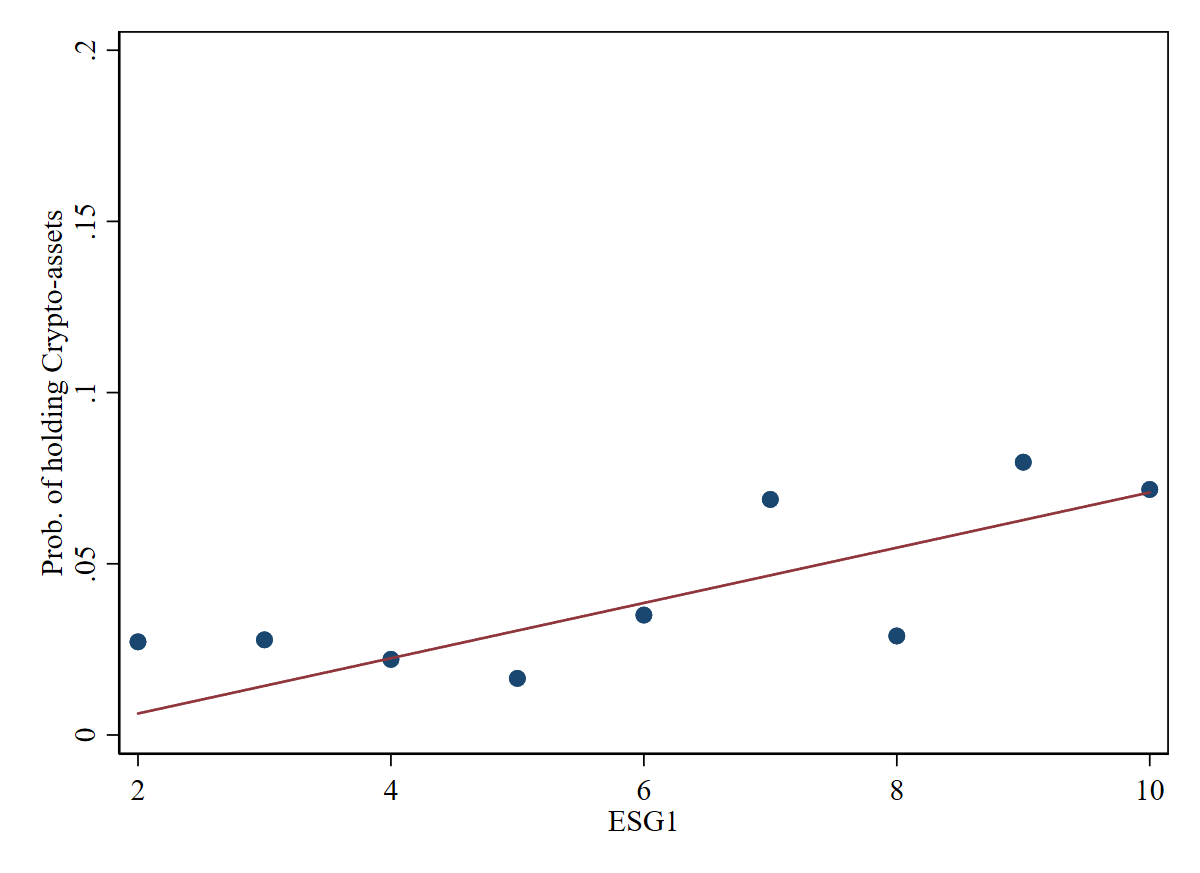}}
\subfloat[Pr. of holding crypto-assets and ESG2 score]{\includegraphics[width=8cm,keepaspectratio]{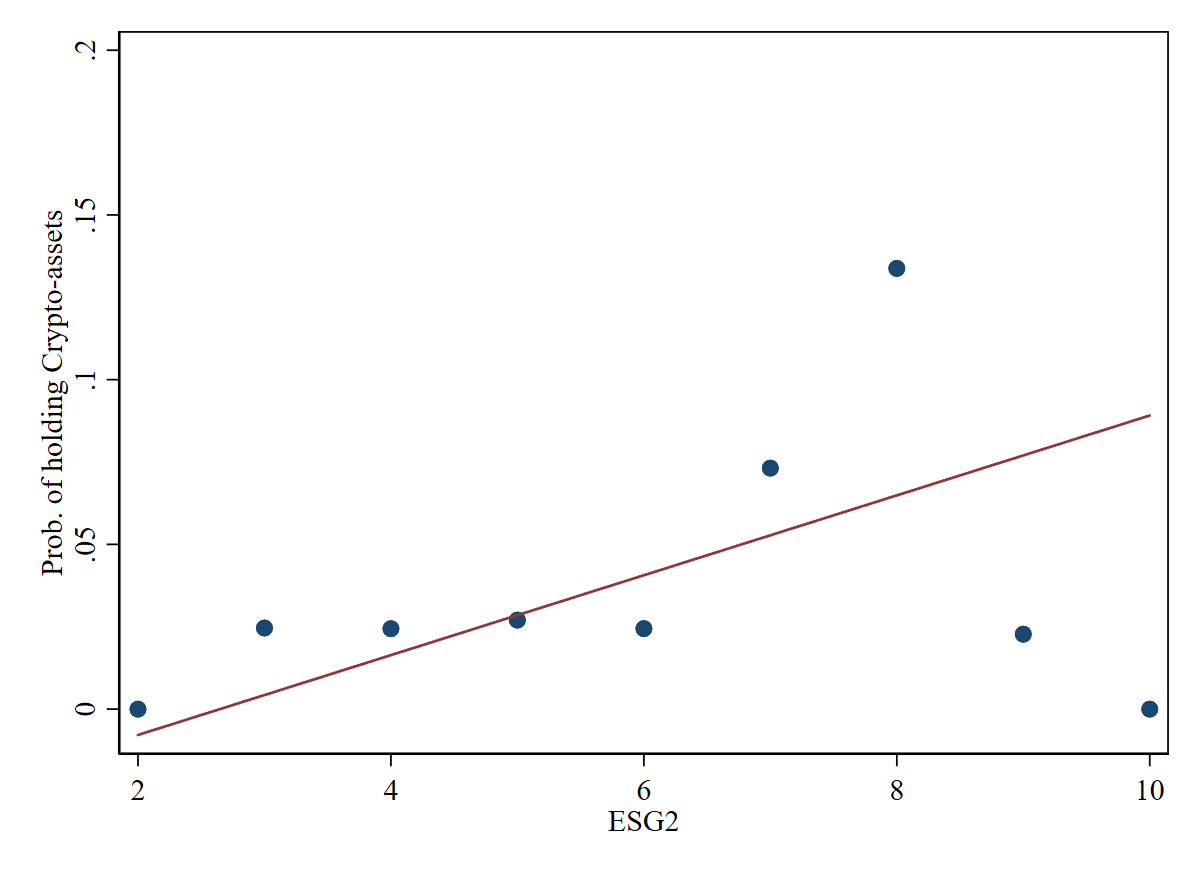}} \\
\subfloat[Pr. of holding bonds and ESG1 score]{\includegraphics[width=8cm,keepaspectratio]{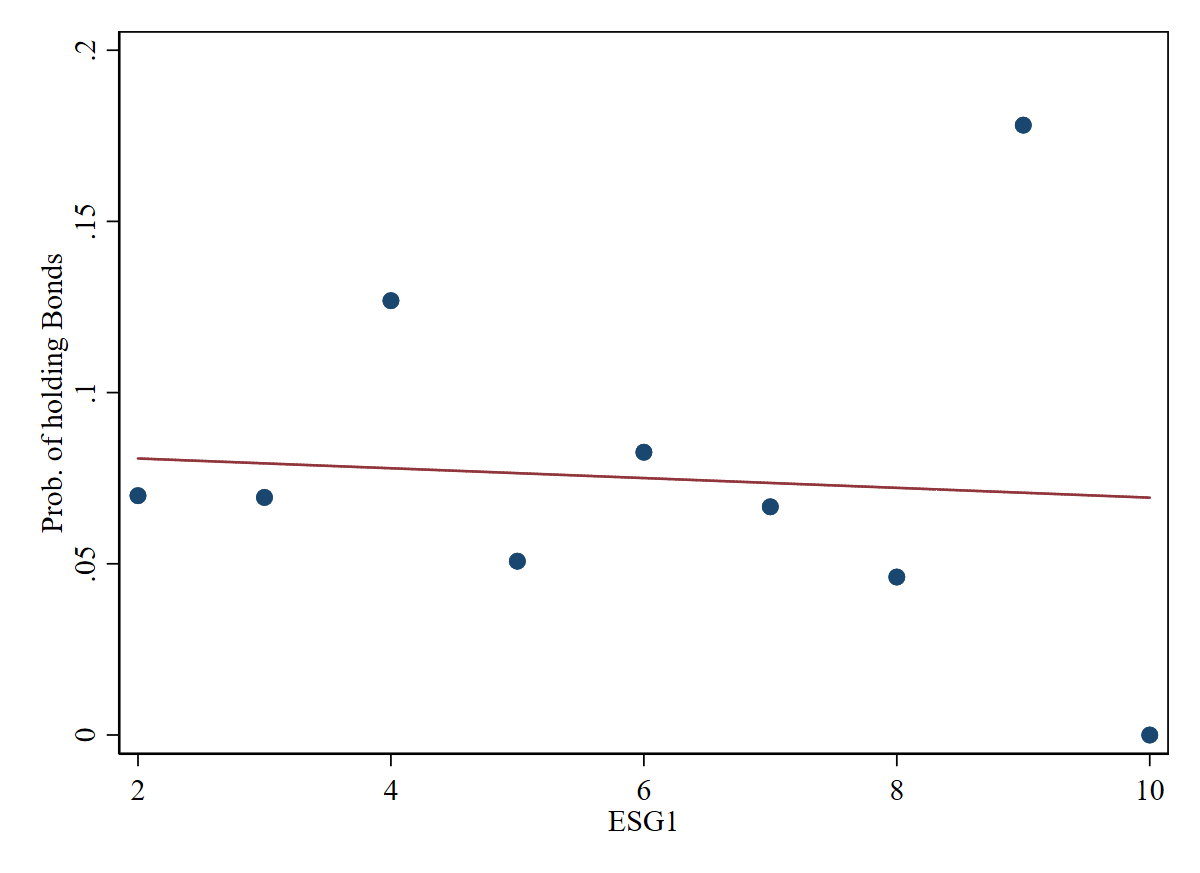}}
\subfloat[Pr. of holding bonds and ESG2 score]{\includegraphics[width=8cm,keepaspectratio]{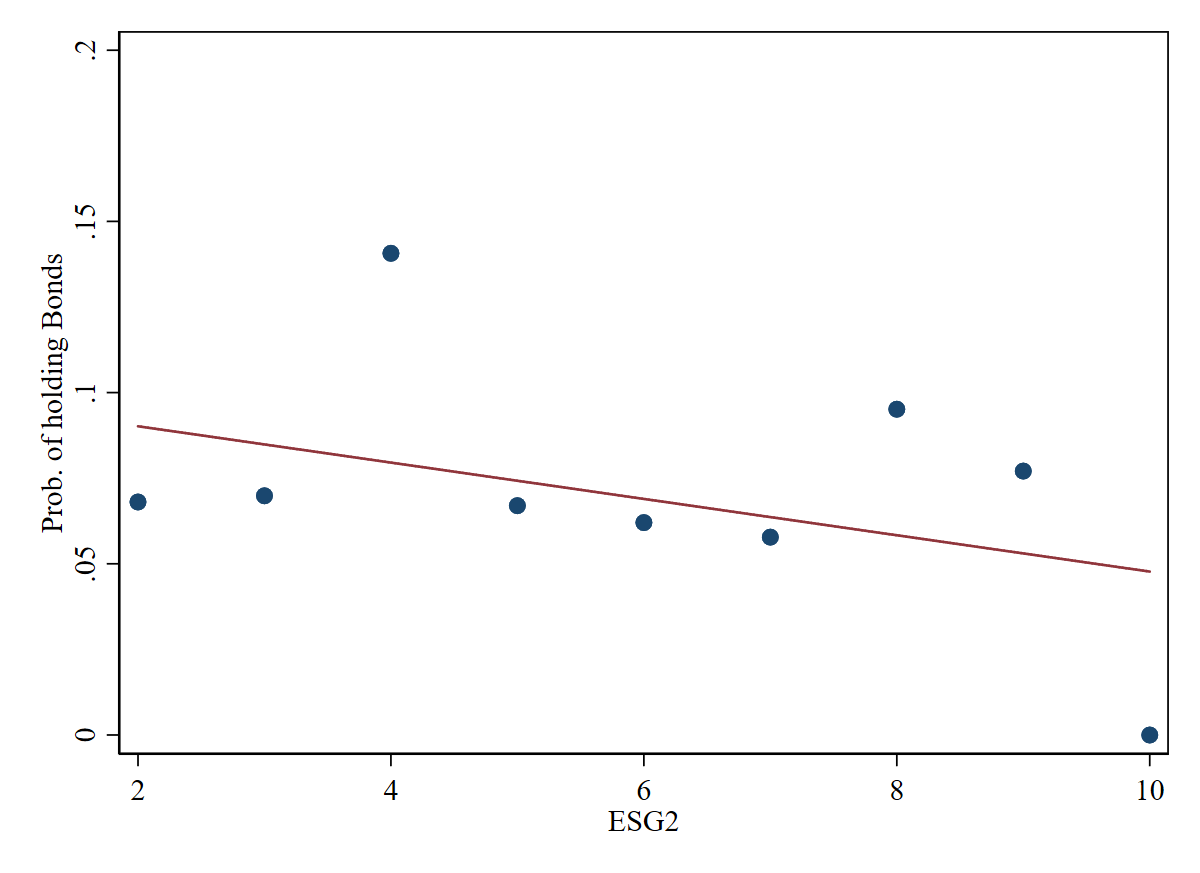}} \\
\subfloat[Pr. of holding shares and ESG1 score]{\includegraphics[width=8cm,keepaspectratio]{graphs/esg2a.png}}
\subfloat[Pr. of holding shares and ESG2 score]{\includegraphics[width=8cm,keepaspectratio]{graphs/esg2b.png}} \\

    \end{center}
\footnotesize{Note: This graph shows binned scatter plots (i.e. reduced form scatter plot) of environmental-social attitudes and holdings of different assets. The probability to hold a certain asset is shown on the vertical axis, while the ESG scores are shown on the horizontal axis.}
\newline \footnotesize{Source: ASFL 2019}
\end{figure}

To gain further insights about the underlying ASFL data, we correlate the computed ESG1 and ESG2 scores with the probability of holding various financial assets: crypto-assets, bonds and shares by means of binned scatter plots (Figure \ref{figure:fig2}). A nuanced and somewhat unexpected pattern emerges: while we observe no relationship between environmental and social attitudes and the probability to own bonds or shares, the relationship is positive and statistically significant for crypto-assets.

\vspace{0.25cm}
\begin{center}
[Figure \ref{figure:fig2} around here]
\end{center}
\vspace{0.25cm}

\section{Estimation approach}
\label{sec:section3}

Our objective is to estimate the relationship between stated investors' ESG preferences and the probability that individuals hold crypto-assets (non-pecuniary effect hypothesis), which we compare to traditional financial asset holdings. In particular, we estimate a linear probability model (LPM) by means of OLS separately for each of the three asset classes (crypto-assets, bonds, shares) using the ASFL data:
\begin{equation}
    Ownership_{ik} = \alpha + \beta_j ESG_{ij} + \gamma X_i + \delta Z_i + \varepsilon_i
\end{equation}
where $Ownership_{ik}$ indicates whether $i$-th individual owns $k$ financial asset, with $k = crypto-assets, bonds, shares$. $ESG_{ij}$ are $i$-th individual's preferences for environmental and social issues, for $j = E, S1, S2, ESG1, ESG2$ (see Table \ref{table:tablea1} in Appendix). $X_i$ represents a set of control variables relevant for individual $i$'s investment decisions, such as age, gender, education, financial literacy, risk aversion, income, etc. To absorb time-invariant cross-sectional variation e.g., in informal institutions, social norms across Austrian provinces, we include regional fixed effects, $Z_i$, in all regressions.

The fact that an individual chooses a certain portfolio allocation might itself affect ESG preferences via different channels such as reading about related developments, being in contact with an investment fund manager or being identified and targeted as a specific consumer for reasons of marketing, though we try to minimise such omitted variable bias by saturating the regression model with economically-relevant covariates related to higher education and financial literacy.

Despite the useful guidance of accumulated evidence from previous studies, it is impossible to know if all important variables have been included. Hence the concern of the ESG endogeneity remains. To address remaining confounders related to potentially endogenous ESG preferences, we use an instrumental variables (IV) approach. To deal with potential endogeneity in the absence of instruments for a standard IV approach, we employ an alternative identification strategy proposed by \cite{lewbel2012}\footnote{The use of this estimation technique is increasingly popular in the household finance literature \citep[e.g.][]{bannier2018,deuflhard2019}. Practical application of this estimation procedure is detailed in \cite{baum2019}.}. It exploits variation on higher moment conditions of the error distribution from the first stage regression of the likely endogenous covariate on (a subset of) other covariates in the model. Identification is achieved by constructing regressors that are uncorrelated with the product of heteroskedastic errors, which is a feature of our data where error correlations are due to an unobserved common factor.

\section{Results}
\label{sec:section4}

\subsection{Main results}

Our baseline model specifications of equation (1) - M1 and M2 - consider alternative composite ESG variables alongside the above detailed explanatory variables. The estimation results employing baseline OLS and \cite{lewbel2012} approach (correcting for potential endogeneity of the ESG) for crypto-assets, bonds, and shares are displayed in Table \ref{table:table2}. For a comparison with baseline results, we estimate additional 4 OLS specifications of equation (1) in order to account for potential multi-collinearity between the explanatory variables and to check the robustness of the estimated coefficients. Models 3 and 4 consider ESG variables individually alongside the relevant socio-economic explanatory variables. Models 5 and 6 are similar to Models 3 and 4 except that they also include financial literacy and financial self-confidence. The estimated OLS results are reported in Table \ref{table:table3}, Table \ref{table:table4} and Table \ref{table:table5} for crypto-assets, bonds, and shares, respectively.

A striking key result is that the non-pecuniary effect hypothesis cannot be rejected based on the ASFL data: ESG-consciousness of investors has a statistically significant impact on individual portfolio exposure to crypto-assets. We observe this positive significant effect of ESG preferences on crypto-asset portfolio composition for both model specifications in the IV estimations in Table \ref{table:table2}. This novel result is also confirmed across most OLS specifications in Table \ref{table:table3}: in M1, M2, M4 and M6. Contrary to a typical crypto-asset perception generated by news media with respect to their ESG footprint, our results indicate that retail investors with stronger ESG preferences invest more likely in crypto-assets than their less-ESG-conscious peers.

Turning to augmented OLS models, they provide an additional specification and robustness checks by confirming that environmental attitudes have a stronger impact on crypto-assets holdings than social attitudes of investors. Further, composite ESG indicators tend to be more statistically significant than individual ESG variables. This result is also confirmed by IV estimates reported in Table \ref{table:table2} where all ESG coefficients are statistically significant and their magnitude is significantly greater than in OLS models.

The results in Table \ref{table:table2} and Table \ref{table:table3} further show that investment in crypto-assets varies by how risk averse investors are in their portfolio choices, by investor's financial literacy and age. Financially better educated and more risk-taking investors are more likely to invest in crypto-assets. Regarding age, older individuals are less likely to invest in crypto-assets - as expected.

These results are in line with the previous literature \citep[e.g.][]{krueger2020}, as investors receive imperfect signals about the crypto-asset ESG-footprint, which usually come from public sources such as news media or from their own idiosyncratic observations. Both risk and ambiguity lead to a cautious investor behaviour and an uncertainty premia in asset markets; learning under risk and ambiguity generates asymmetric responses to ESG-news. ESG preferences affect investment decisions because they serve as a proxy for value-relevant information or risk, they enhance performance or reduce risk.

As a benchmark, we compare the crypto-asset holding probabilities with holding probabilities of traditional risky assets, namely bonds and shares in Table \ref{table:table2}. While the estimated relationship between ESG preferences and crypto holdings is positive and statistically significant, we find no such a statistically significant relationship between ESG preferences and the probability to hold bonds or shares. OLS estimates in Table \ref{table:table4} and Table \ref{table:table5} confirm these findings. This result finds strong support in the recent empirical literature on ESG investing. For example, \cite{anderson2021} have not found any statistically significant relationship between individuals' ESG attitudes and ownership of pro-environment portfolios (green bonds, stocks, and pension funds) in a sample of Swedish households. This implies that in our estimations,
which are based on the AFLS data that do not identify separately ESG stocks and non-ESG assets, the relationship between ESG preferences and the probability to hold traditional assets are even less likely to be present if the findings of \cite{anderson2021} were generalisable for Austria.

\vspace{0.25cm}

\begin{center}
[Table \ref{table:table2} around here]
\end{center}

\begin{center}
[Table \ref{table:table3} around here]
\end{center}

\begin{center}
[Table \ref{table:table4} around here]
\end{center}

\begin{center}
[Table \ref{table:table5} around here]
\end{center}
\vspace{0.25cm}

For the household finance literature that studies determinants of portfolio holdings, our results add a further piece of evidence that non-pecuniary effects indeed matter in explaining cross-sectional differences in investment decisions; whereby the association between ESG and crypto-assets is stronger compared to traditional risky assets like bonds and shares.

\begin{figure}[!h]
\caption{Evolution of Proof-of-Stake market share over time}
    \begin{center}
    \label{figure:fig3}
    {\includegraphics[width=17cm,keepaspectratio]{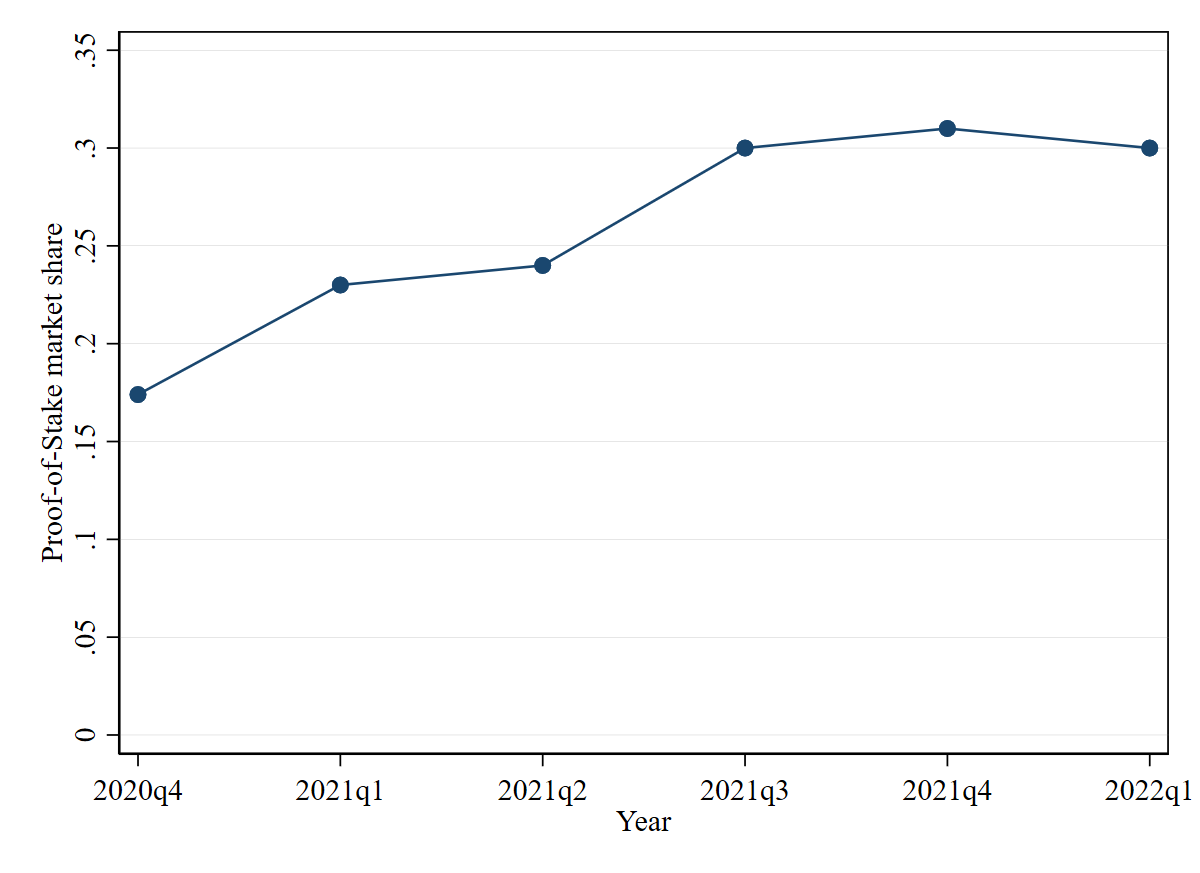}}
    \end{center}
\footnotesize{Source: Based on data from https://staking.staked.us/state-of-staking}
\end{figure}

There are two recently documented facts that support our findings. On the supply side, many crypto-assets have a low (even zero) ESG footprint, including the PoS class of consensus mechanisms and the usage of renewable energy sources for mining. Moreover, the share of sustainable crypto-assets is increasing continuously. For example, since its introduction 10 years ago, the PoS market share has reached 30\% in 2021 (see Figure \ref{figure:fig3}), its energy consumption is 99.5\% lower compared to PoW. Even the PoW-based blockchains are increasingly ``decarbonise'' by being mined using renewable energy sources like solar, hydro or wind power, e.g. in Iceland, Norway (Crypto Climate Accord). Indeed, the wide range of cross-sectional distributions within the crypto-asset class allows forward-looking ESG-conscious investors to match closely their preferences, subjective beliefs, ESG performance, risk aversion, etc. \citep{saleh2021}.

\vspace{0.25cm}
\begin{center}
[Figure \ref{figure:fig3} around here]
\end{center}
\vspace{0.25cm}

On the demand side, the underlying ASFL data cover individual investors, who compared to large corporate crypto-asset holders tend to have stronger non-traditional (imperfectly rational) preferences for the portfolio ESG footprint \citep{mustafa2022}. Further, retail crypto investors are young, above-average educated, and financially more literate compared to the general population, and younger cohorts tend to have stronger environmental concerns than older cohorts \citep{stix2021,fujiki2021}. In the era of digital disruption, which is continuing to fragment the crypto-asset market, and the growing number of investment tools hitting the market empower small individual ESG-conscious investors by giving them the ability to take control of whether their money is being invested for good in the world.

For the crypto-asset literature, the evidence we provide is supportive of crypto-asset-related environmental concerns (e.g. high energy consumption in the PoW mining) being of first-order for crypto holdings, whereas governance and social issues (e.g. decentralised governance and financial inclusion) of second-order. That is, ESG-conscious investors tend to invest more often in crypto-assets even though in the general crypto-asset class there are also cryptocurrencies with adverse environmental effects for example due to high energy consumption. The existence of alternative less energy-intensive consensus mechanisms, e.g., the PoS is much less energy intensive than PoW, and the usage of renewable energy sources for mining may explain our results \citep{platt2021}. We find less support for a causal relationship between non-pecuniary effects related to social and governance preferences in Austrian individual investor portfolio exposure to crypto-assets.

\subsection{Further analysis and robustness}

We estimate several additional models serving as robustness checks, for diagnostic purposes and transparency. First, we check if the coefficients remain stable after accounting for possible non-linearities in effects of age and income. The results suggest that even considering the non-linear quadratic terms do not alter our main set of estimated ESG effects (see Table \ref{table:tablea2} in Appendix). Second, given the binary nature of our dependent variable (ownership of crypto-assets), we estimate a set of probit regressions (results shown in Table \ref{table:tablea3}) to check the robustness of our main OLS estimates presented in Table \ref{table:table2} through Table \ref{table:table5}. We can see that probit marginal effects are somewhat smaller compared to OLS, but still similar in magnitude. Finally, given the rare occurrence of the crypto-assets owners (around 3\% of the sample), simple OLS or probit estimates might suffer from bias as suggested by \cite{king2001}. Therefore, we have re-estimated our main OLS and probit models by means of a rare-events logit model\footnote{To estimate the rare-events logit model, we use the Stata estimation command `relogit' implemented by \cite{tomz2021}}. We report estimated results from three estimation procedures next to each other in Table \ref{table:tablea3} in Appendix and can see that the OLS/LPM estimates are quite close to the marginal effects obtained from the estimated coefficients for rare-events logit model. This supports the OLS estimation approach also in the 2SLS IV framework.

\section{Conclusions}
\label{sec:section5}

We studied the relevance of non-pecuniary effects in driving cross-sectional differences in investment decision. In particular, we examined the relationship between ESG preferences and holdings of crypto-assets; and compared how the investors' ESG preferences effect on investment decisions differ between crypto-assets and traditional financial assets.

Our results suggest that on average individuals with stronger ESG preferences tend to invest more frequently in crypto-assets than less ESG-conscious investors. Second, the association between environmental attitudes and crypto investments is of first-order, whereas social and governmental attitudes do not determine the portfolio exposure to crypto-assets of ESG-conscious investors. Our paper delivers a novel evidence regarding the ESG preferences of individual investors exhibiting a subjective belief dynamics - in line with the household finance literature finding that a priori stated socially ``desirable'' preferences do not always match the preferences revealed in the portfolio choice \citep{anderson2021}. Contrary to a typical crypto-asset perception generated by news media with respect to their ESG footprint, our results indicate that retail investors with stronger ESG preferences invest more likely in crypto-assets than their less-ESG-conscious peers. However, there are also other potential reasons why such a result could actually be in line with consistent preferences with regard to communication and actual portfolio choice.

On the supply side, many new generation crypto-assets have an extremely low ESG footprint, including the PoS class of consensus mechanisms and the usage of green renewable energy sources in mining. Moreover, the share of sustainable crypto-assets is continuously increasing. The wide range of distributions within the crypto-asset class allows forward-looking ESG-conscious investors to match closely their preferences, subjective beliefs, ESG performance, risk aversion, etc. On the demand side, the digital disruption which is continuing to fragment the crypto-asset market, and the growing number of investment tools available on the market attracts small individual ESG-conscious investors giving them the ability to take control of whether their money is being used in line with their ESG preferences. Indeed, the individual investors, who compared to large corporate crypto-asset holders tend to exhibit stronger non-pecuniary preferences for their portfolio ESG footprint, are young, above-average educated, and financially more literate compared to the general population.

These findings suggest that non-pecuniary effects of crypto-investors captured via social and ethical/moral preferences should be (and are already) taken into consideration, when designing new digital currencies, e.g. as is under discussion by a number of central banks. Second, the value added of the inclusion of separate items and more detailed information on crypto-assets and other alternative financial instruments in standard finance and wealth surveys becomes evident. Our results also highlight the need to collect detailed information on investor's beliefs and attitudes within the household portfolio context, beyond the standard socio-economic variables to better understand individual investment decisions.

While paper delivered first insights, we strongly believe that more research is needed using larger household finance datasets which allow for a more detailed and comprehensive socio-economic analysis of the relationship of ESG preferences and portfolio choice with regard to crypto-assets. For this reason we call for an inclusion of crypto-asset questions into standard household finance surveys such as the Survey of Consumer Finances (US), The Wealth and Asset Survey (UK) or the Household Finance and Consumption Survey (Continental Europe). Only survey data which includes extensive and intensive margins of crypto-asset holdings along with the rest of the household balance sheet as well as a large number of socio-economic characteristics and preferences will allow to create a deeper understanding of portfolio choice with regard to crypto-assets. Such a micro-evidence-based understanding is urgently needed given the quick rise of crypto-assets especially among the younger investor cohorts, not only for potential regulation purposes but also to monitor the financial behaviour of households and potential risks created for the financial stability.

\clearpage
\bibliographystyle{ecca}
\bibliography{references}

\begin{thebibliography}{34}
\providecommand{\natexlab}[1]{#1}

\bibitem[{Anderson and Robinson(2021)}]{anderson2021}
\textsc{Anderson, A.} and \textsc{Robinson, D.~T.} (2021). Financial literacy
  in the age of green investment. \textit{Review of Finance}, \textbf{63}~(12),
  1572--3097.

\bibitem[{Bannier \textit{et~al.}(2019)Bannier, Meyll, R{\"o}der and
  Walter}]{bannier2019}
\textsc{Bannier, C.}, \textsc{Meyll, T.}, \textsc{R{\"o}der, F.} and
  \textsc{Walter, A.} (2019). The gender gap in ‘bitcoin literacy’.
  \textit{Journal of Behavioral and Experimental Finance}, \textbf{22},
  129--134.

\bibitem[{Bannier and Schwarz(2018)}]{bannier2018}
\textsc{Bannier, C.~E.} and \textsc{Schwarz, M.} (2018). Gender-and
  education-related effects of financial literacy and confidence on financial
  wealth. \textit{Journal of Economic Psychology}, \textbf{67}, 66--86.

\bibitem[{Barone and Masciandaro(2019)}]{barone2019}
\textsc{Barone, R.} and \textsc{Masciandaro, D.} (2019). Cryptocurrency or
  usury? crime and alternative money laundering techniques. \textit{European
  Journal of Law and Economics}, \textbf{47}~(2), 233--254.

\bibitem[{Baum and Lewbel(2019)}]{baum2019}
\textsc{Baum, C.~F.} and \textsc{Lewbel, A.} (2019). Advice on using
  heteroskedasticity-based identification. \textit{Stata Journal},
  \textbf{19}~(4), 757--767.

\bibitem[{Bouri \textit{et~al.}(2019)Bouri, Gupta and Roubaud}]{bouri2019}
\textsc{Bouri, E.}, \textsc{Gupta, R.} and \textsc{Roubaud, D.} (2019). Herding
  behaviour in cryptocurrencies. \textit{Finance Research Letters},
  \textbf{29}, 216--221.

\bibitem[{Chapron(2017)}]{chapron2017}
\textsc{Chapron, G.} (2017). The environment needs cryptogovernance.
  \textit{Nature}, \textbf{545}~(7655), 403--405.

\bibitem[{Chen \textit{et~al.}(2020)Chen, Hansen and Hansen}]{chen2020}
\textsc{Chen, X.}, \textsc{Hansen, L.~P.} and \textsc{Hansen, P.~G.} (2020).
  Robust identification of investor beliefs. \textit{Proceedings of the
  National Academy of Sciences}, \textbf{117}~(52), 33130--33140.

\bibitem[{Ciaian \textit{et~al.}(2016)Ciaian, Rajcaniova and
  Kancs}]{ciaian2016}
\textsc{Ciaian, P.}, \textsc{Rajcaniova, M.} and \textsc{Kancs, d.} (2016). The
  digital agenda of virtual currencies: Can bitcoin become a global currency?
  \textit{Information Systems and e-Business Management}, \textbf{14}~(4),
  883--919.

\bibitem[{Deuflhard \textit{et~al.}(2019)Deuflhard, Georgarakos and
  Inderst}]{deuflhard2019}
\textsc{Deuflhard, F.}, \textsc{Georgarakos, D.} and \textsc{Inderst, R.}
  (2019). Financial literacy and savings account returns. \textit{Journal of
  the European Economic Association}, \textbf{17}~(1), 131--164.

\bibitem[{Dilek and Furuncu(2019)}]{dilek2019}
\textsc{Dilek, {\c{S}}.} and \textsc{Furuncu, Y.} (2019). Bitcoin mining and
  its environmental effects. \textit{Journal of Economics and Administrative
  Sciences}, \textbf{33}~(1), 91--106.

\bibitem[{Duarte \textit{et~al.}(2021)Duarte, Fonseca, Goodman and
  Parker}]{duarte2021}
\textsc{Duarte, V.}, \textsc{Fonseca, J.}, \textsc{Goodman, A.~S.} and
  \textsc{Parker, J.~A.} (2021). \textit{Simple Allocation Rules and Optimal
  Portfolio Choice Over the Lifecycle}. Working Paper 29559, National Bureau of
  Economic Research.

\bibitem[{Ehrlich and Yin(2022)}]{ehrlich2022}
\textsc{Ehrlich, I.} and \textsc{Yin, Y.} (2022). \textit{A Cross-Country
  Comparison of Old Age Financial Readiness in Asian Countries vs. the United
  States: The Case of Japan and the Republic of Korea}. Working Paper 29649,
  National Bureau of Economic Research.

\bibitem[{Fessler \textit{et~al.}(2020)Fessler, Jelovsek, Silgoner
  \textit{et~al.}}]{fessler2020}
\textsc{Fessler, P.}, \textsc{Jelovsek, M.}, \textsc{Silgoner, M.}
  \textit{et~al.} (2020). Financial literacy in austria--focus on millennials.
  \textit{Monetary Policy and the Economy}, \textbf{3}, 21--38.

\bibitem[{Fujiki(2021)}]{fujiki2021}
\textsc{Fujiki, H.} (2021). Crypto asset ownership, financial literacy, and
  investment experience. \textit{Applied Economics}, \textbf{53}~(39),
  4560--4581.

\bibitem[{Jiang \textit{et~al.}(2021)Jiang, Peng and Yan}]{jiang2021}
\textsc{Jiang, Z.}, \textsc{Peng, C.} and \textsc{Yan, H.} (2021). Personality
  differences and investment decision-making. \textit{SSRN 3580364}.

\bibitem[{King and Zeng(2001)}]{king2001}
\textsc{King, G.} and \textsc{Zeng, L.} (2001). Logistic regression in rare
  events data. \textit{Political Analysis}, \textbf{9}~(2), 137--163.

\bibitem[{Kohler and Pizzol(2019)}]{kohler2019}
\textsc{Kohler, S.} and \textsc{Pizzol, M.} (2019). Life cycle assessment of
  bitcoin mining. \textit{Environmental Science \& Technology},
  \textbf{53}~(23), 13598--13606.

\bibitem[{Krueger \textit{et~al.}(2020)Krueger, Sautner and
  Starks}]{krueger2020}
\textsc{Krueger, P.}, \textsc{Sautner, Z.} and \textsc{Starks, L.~T.} (2020).
  The importance of climate risks for institutional investors. \textit{Review
  of Financial Studies}, \textbf{33}~(3), 1067--1111.

\bibitem[{Lewbel(2012)}]{lewbel2012}
\textsc{Lewbel, A.} (2012). Using heteroscedasticity to identify and estimate
  mismeasured and endogenous regressor models. \textit{Journal of Business \&
  Economic Statistics}, \textbf{30}~(1), 67--80.

\bibitem[{Liu and Peifer(2022)}]{liu2022}
\textsc{Liu, J.} and \textsc{Peifer, J.~L.} (2022). A moral foundations framing
  approach: Retail investors’ investment intention in ethical mutual funds.
  \textit{Business \& Society}, pp. 1--27.

\bibitem[{Lusardi and Mitchell(2014)}]{lusardi2014}
\textsc{Lusardi, A.} and \textsc{Mitchell, O.~S.} (2014). The economic
  importance of financial literacy: Theory and evidence. \textit{Journal of
  Economic Literature}, \textbf{52}~(1), 5--44.

\bibitem[{Mustafa \textit{et~al.}(2022)Mustafa, Lodh, Nandy and
  Kumar}]{mustafa2022}
\textsc{Mustafa, F.}, \textsc{Lodh, S.}, \textsc{Nandy, M.} and \textsc{Kumar,
  V.} (2022). Coupling of cryptocurrency trading with the sustainable
  environmental goals: Is it on the cards? \textit{Business Strategy and the
  Environment}, \textbf{31}~(3), 1152--1168.

\bibitem[{OECD(2018)}]{oecd2018}
\textsc{OECD} (2018). \textit{OECD/INFE toolkit for measuring financial
  literacy and financial inclusion}. Tech. rep., OECD Publishing, Paris.

\bibitem[{Parmentola \textit{et~al.}(2022)Parmentola, Petrillo, Tutore and
  De~Felice}]{parmentola2022}
\textsc{Parmentola, A.}, \textsc{Petrillo, A.}, \textsc{Tutore, I.} and
  \textsc{De~Felice, F.} (2022). Is blockchain able to enhance environmental
  sustainability? \textit{Business Strategy and the Environment},
  \textbf{31}~(1), 194--217.

\bibitem[{Pastor \textit{et~al.}(2021a)Pastor, Stambaugh and
  Taylor}]{pastor2021a}
\textsc{Pastor, L.}, \textsc{Stambaugh, R.~F.} and \textsc{Taylor, L.~A.}
  (2021a). \textit{Dissecting Green Returns}. Working Paper 28940, National
  Bureau of Economic Research.

\bibitem[{Pastor \textit{et~al.}(2021b)Pastor, Stambaugh and
  Taylor}]{pastor2021b}
\textsc{---}, \textsc{---} and \textsc{---} (2021b). Sustainable investing in
  equilibrium. \textit{Journal of Financial Economics}, \textbf{142}~(2),
  550--571.

\bibitem[{Platt \textit{et~al.}(2021)Platt, Sedlmeir, Platt, Tasca, Xu, Vadgama
  and Ibanez}]{platt2021}
\textsc{Platt, M.}, \textsc{Sedlmeir, J.}, \textsc{Platt, D.}, \textsc{Tasca,
  P.}, \textsc{Xu, J.}, \textsc{Vadgama, N.} and \textsc{Ibanez, J.~I.} (2021).
  Energy footprint of blockchain consensus mechanisms beyond proof-of-work.
  \textit{arXiv:2109.03667}.

\bibitem[{Richman \textit{et~al.}(2021)Richman, Frankovitz and
  McDonald}]{richman2021}
\textsc{Richman, B.}, \textsc{Frankovitz, N.} and \textsc{McDonald, L.} (2021).
  \textit{Crypto \& ESG White Paper}. Tech. rep., Sarson Funds.

\bibitem[{Saleh(2021)}]{saleh2021}
\textsc{Saleh, F.} (2021). Blockchain without waste: Proof-of-stake.
  \textit{Review of Financial Studies}, \textbf{34}~(3), 1156--1190.

\bibitem[{Stix(2021)}]{stix2021}
\textsc{Stix, H.} (2021). Ownership and purchase intention of crypto-assets:
  survey results. \textit{Empirica}, \textbf{48}~(1), 65--99.

\bibitem[{Teichmann and Falker(2021)}]{teichmann2021}
\textsc{Teichmann, F. M.~J.} and \textsc{Falker, M.-C.} (2021). Money
  laundering via cryptocurrencies - potential solutions from liechtenstein.
  \textit{Journal of Money Laundering Control}, \textbf{24}~(1), 91--101.

\bibitem[{Tomz \textit{et~al.}(2021)Tomz, King and Zeng}]{tomz2021}
\textsc{Tomz, M.}, \textsc{King, G.} and \textsc{Zeng, L.} (2021).
  \textit{{RELOGIT: Stata module to perform Rare Event Logistic Regression}}.
  Tech. rep., Statistical Software Components, Boston College Department of
  Economics.

\bibitem[{Xi \textit{et~al.}(2020)Xi, O’Brien and Irannezhad}]{xi2020}
\textsc{Xi, D.}, \textsc{O’Brien, T.~I.} and \textsc{Irannezhad, E.} (2020).
  Investigating the investment behaviors in cryptocurrency. \textit{Journal of
  Alternative Investments}, \textbf{23}~(2), 141--160.

\end{thebibliography}

\clearpage
\section*{Figures and Tables}


\clearpage
\newpage


\clearpage
\newpage
\begin{table}[!h]
\caption{Descriptive statistics}
\label{table:table1}
\centering
\resizebox{16cm}{!}{
\begin{tabular}{l*{5}{c}}
\hline\hline
Variable	                            &N    &Mean	&SD	    &Min	  &Max \\
\hline
Crypto-assets ownership	                &1402	&0.03	&0.18	&0	&1 \\
Bonds ownership	                        &1398	&0.07	&0.25	&0	&1 \\
Stocks/shares ownership	                &1404	&0.11	&0.31	&0	&1 \\
Preferences for enviro. issues (E)   	&1274	&3.72	&1.15	&1	&5 \\
Preferences for social issues (S1)   	&1198	&2.18	&1.01	&1	&5 \\
Preferences for social issues (S2)	    &1363	&2.03	&0.97	&1	&5 \\
ESG1 (E + S1)	                        &1126	&5.83	&1.52	&2	&10 \\
ESG2 (E + S2)	                        &1250	&5.75	&1.42	&2	&10 \\
Objective fin. literacy	                &1418	&5.32	&1.64	&0	&7 \\
Confidence in own fin. knowledge	    &1382	&3.27	&0.98	&1	&5 \\
Risk attitude score                 	&1418	&1.57	&0.82	&1	&4 \\
Primary education	                    &1382	&0.14	&0.35	&0	&1 \\
Secondary education	                    &1382	&0.76	&0.43	&0	&1 \\
Tertiary education	                    &1382	&0.10	&0.30	&0	&1 \\
Gender: female	                        &1418	&0.52	&0.50	&0	&1 \\
Age	                                    &1418	&49.08	&18.20	&16	&97 \\
Individual monthly income	            &1188	&1642.25 &812.35	&0	&5250 \\
\hline\hline
\end{tabular}}
\begin{tablenotes}
   \item{\footnotesize {Note: Summary statistics computed using survey weights. There are three main regions (Region of East Austria, Region of South Austria, and Region of West Austria), which are equally represented in the survey. \\
Source: ASFL 2019.}}
   \end{tablenotes}
\end{table}


\clearpage
\newpage
\begin{landscape}
\begin{table}[!h]\centering
 \caption{Results on ESG preferences for financial assets (OLS and IV)}
 \label{table:table2}
 \resizebox{22.5cm}{!}{
 \begin{tabular}{lcccccccccccccc}
\hline\hline
&\multicolumn{4}{c}{Crypto-assets} & &\multicolumn{4}{c}{Bonds}	  & &\multicolumn{4}{c}{Shares} \\ \cline{2-5} \cline{7-10} \cline{12-15}
	                               &(M1)  	&(M1)	 &(M2) &(M2) &	&(M1)	&(M1) &(M2)  &(M2) & &(M1) &(M1)	&(M2)	&(M2) \\
	                               &OLS  	&IV	 &OLS &IV &	&OLS  	&IV	 &OLS &IV & &OLS  	&IV	 &OLS &IV \\
\hline
ESG1	                         &0.008*	&0.026**	&&&			&0.004	&0.010		&&&		&-0.006	&-0.010	\\	
	                             &(0.005)	&(0.012)	&&&			&(0.005) &(0.014)	&&&			&(0.009)	&(0.029) \\		
ESG2			                   &&            &0.010**	&0.028*	 &&&		&-0.003	&-0.001	 &&&	&-0.012	&-0.024 \\
			                        &&            &(0.005)	&(0.015) &&&		&(0.006) &(0.015) &&&	&(0.009) &(0.034) \\
Objective fin. literacy 	       &0.013**	&0.014**	&0.016***	&0.017***	&	&0.001	&0.001	&0.002	&0.002	&	&0.016*	&0.016*	&0.016**	&0.015* \\
	                               &(0.006)	&(0.006)	&(0.005)	&(0.006) &	&(0.006)	&(0.006)	&(0.005)	&(0.005) &	&(0.009)	&(0.009)	&(0.008)	&(0.009) \\
Confidence in own fin. knowledge 	&0.013*	&0.013*	&0.013*	&0.013*	&	&0.013	&0.013	&0.012	&0.012	&	&0.019	&0.020	&0.022**	&0.022** \\
	                               &(0.007)	&(0.007)	&(0.007)	&(0.007) &		&(0.009)	&(0.009)	&(0.008)	&(0.008)	&	&(0.012)	&(0.012)	&(0.011)	&(0.011) \\
Risk attitude score 	          &0.056***	&0.054***	&0.058***	&0.053*** &	&0.044*** &0.044***	&0.049***	&0.048*** &	&0.128***	&0.129***	&0.127***	&0.130*** \\
	                               &(0.014)	&(0.014)	&(0.014)	&(0.013) &		&(0.012)	&(0.012)	&(0.012)	&(0.013) &		&(0.018)	&(0.018)	&(0.017)	&(0.019) \\
Secondary education 	          &0.004	&0.004	&0.003	&0.002	&	&0.032*	&0.032*	&0.024	&0.024	&	&0.042	&0.042	&0.027	&0.028 \\
	                           &(0.016)	&(0.016)	&(0.014)	&(0.015) &		&(0.018)	&(0.018)	&(0.016)	&(0.016)	&	&(0.030)	&(0.030)	&(0.026)	&(0.025) \\
Tertiary education 	                 &-0.025	&-0.023	&-0.029	&-0.024	&	&0.075*	&0.076*	&0.064	&0.064	&	&0.092*	&0.091*	&0.062	&0.059 \\
	                                &(0.026)	&(0.026)	&(0.024)	&(0.025)	&	&(0.043)	&(0.043)	&(0.041)	&(0.040) &	&(0.055)	&(0.054)	&(0.050)	&(0.047) \\
Individual monthly income 	          &-0.000	&-0.000	&-0.000	&-0.000	&	&0.000***	&0.000***	&0.000***  &0.000***	&	&0.000***	&0.000***	&0.000*** &0.000*** \\
	                                 &(0.000)	&(0.000)	&(0.000)	&(0.000) &		&(0.000)	&(0.000)	&(0.000)	&(0.000) &	&(0.000)	&(0.000)	&(0.000)	&(0.000) \\
Gender: female                  	&-0.008	&-0.003	&0.004	&0.008	&	&-0.004	&-0.003	&-0.017	&-0.016	&	&0.011	&0.010	&-0.005	&-0.007 \\
	                                  &(0.012)	&(0.013)	&(0.013)	&(0.014) &		&(0.016)	&(0.017)	&(0.015)	&(0.015) &		&(0.020)	&(0.021)	&(0.019)	&(0.019) \\
Age 	                             &-0.001**	&-0.001**	&-0.001**	&-0.001** &		&0.002***	&0.002***	&0.002***	&0.002*** &		&0.002***	&0.002***	&0.002***	&0.002*** \\
                                	&(0.000)	&(0.000)	&(0.000)	&(0.000) &		&(0.001)	&(0.001)	&(0.001)	&(0.000) &		&(0.001)	&(0.001)	&(0.001)	&(0.001) \\
Constant                             &-0.155***	&-0.121*	&-0.175***	&-0.129	&	&-0.306***	&0.007	&-0.242***	&0.070	&	&-0.472***	&0.167	&-0.389***	&0.243 \\
	                                 &(0.049)	&(0.070)	&(0.049)	&(0.084) &		&(0.066)	&(0.086)	&(0.061)	&(0.090) &		&(0.111)	&(0.173)	&(0.104)	&(0.199) \\

\hline
Regional fixed effects	          &YES	&YES	&YES	&YES	&	&YES	&YES	&YES	&YES &		&YES	&YES	&YES	&YES \\
R2	                              &0.10	&	&0.11	&&		&0.12	&	&0.12	&		&0.21 &		&0.21 & \\	
N	                              &902	&902	&1000	&1000 &		&904	&904	&998	&998 &		&903	&903	&1000	&1000 \\
F-stat		                      &     &11.43	&	&16.09	&&	&11.59	&	&16.61	&&		&11.59	&	&16.28 \\
Hansen J-test		              &      &8.89	&	&8.49	&&		&9.55	&	&11.95 &&		&12.95	&	&7.79 \\
Hansen J-test (p-value)		      &      &0.45	&	&0.49	&&		&0.39	&	&0.22	&&		&0.16	&	&0.56 \\
Breusch-Pagan-test		          &      &4.27	&	&12.98	&&		&4.12	&	&12.27	&&		&3.97	&	&11.76 \\
Breusch-Pagan-test (p-value)	&	     &0.04	&	&0.00	&&		&0.04	&	&0.00	&&		&0.05	&	&0.00 \\

\hline\hline
\end{tabular}}
\begin{tablenotes}
    \item \footnotesize{Note: Regressions estimated using survey weights. Robust standard errors are reported in parentheses. Dummy variable for ‘Primary education’ category is the reference category of the respective dummy variables set. All RHS covariates (i.e. instruments) in the IV models have been generated according to the \cite{lewbel2012} methodology which is implemented within the Stata ‘ivreg2h’ estimation command \citep{baum2019}. \\
* p<0.1, ** p<0.05, *** p<0.01.}
    \item \footnotesize{Source: Own estimates based on ASFL 2019 data}
    \end{tablenotes}
\end{table}
\end{landscape}


\clearpage
\newpage
\begin{table}[!h]\centering
 \caption{Results on ESG preferences for crypto-assets (OLS)}
 \label{table:table3}
 \resizebox{16cm}{!}{
 \begin{tabular}{lcccccc}
\hline\hline
	                                      &(M1)  	&(M2)	 &(M3)	      &(M4)	&(M5)	&(M6) \\
\hline
Preferences for enviro. issues (E)			& &                &0.004	&0.011**	&0.005 &0.012** \\
			                                 &&                &(0.005)	&(0.005)	&(0.005) &(0.005) \\
Preferences for social issues (S1)		     &&               & 0.011	   & 	&0.013	\\
			                                 &&                & (0.008)   &	&(0.008) \\	
Preferences for social issues (S2)			&&&	                      &0.004	&	&0.007 \\
				                              &&&                    &(0.007)	&	&(0.007) \\
ESG1	                               &0.008*			&&&&& \\		
	                                &(0.005)			&&&&& \\		
ESG2		                          &          &0.010**	 &&&& \\			
		                              &          &(0.005)	 &&&& \\			
Objective fin. literacy            	&0.013**	&0.016***	&&		&0.013**	&0.015*** \\
	                               &(0.006)	&(0.005)		&&	   &(0.006)	   &(0.006) \\
Confidence in own fin. knowledge	&0.013*	&0.013*	 &&		&0.013*	 &0.013* \\
	                               &(0.007)	&(0.007) &&		&(0.007)	&(0.007) \\
Risk attitude score	                 &0.056***	&0.058***	&0.058***	&0.060***	&0.056***	&0.059*** \\
	                                &(0.014)	&(0.014)	&(0.014)	&(0.013)	&(0.014)	&(0.014) \\
Secondary education	               &0.004	&0.003	&0.018	&0.021	&0.004	&0.004 \\
	                                &(0.016)	&(0.014)	&(0.017)	&(0.015)	&(0.016)	&(0.014) \\
Tertiary education	               &-0.025	&-0.029	&0.002	&0.001	&-0.024	&-0.029 \\
	                               &(0.026)	&(0.024)	&(0.024)	&(0.021)	&(0.026)	&(0.024) \\
Individual monthly income	       &-0.000	&-0.000	&-0.000	&-0.000	&-0.000	&-0.000 \\
	                               &(0.000)	&(0.000)	&(0.000)	&(0.000)	&(0.000)	&(0.000) \\
Gender: female	                   &-0.008	&0.004	&-0.014	&-0.002	&-0.008	&0.004 \\
	                               &(0.012)	&(0.013)	&(0.013)	&(0.013)	&(0.012)	&(0.013) \\
Age	                               &-0.001**	&-0.001**	&-0.000	&-0.001**	&-0.001*	&-0.001** \\
	                               &(0.000)	&(0.000)	&(0.000)	&(0.000)	&(0.000)	&(0.000) \\
Constant                        	&-0.155***	&-0.175***	&-0.069*	&-0.080**	&-0.157***	&-0.174*** \\
	                               &(0.049)	&(0.049)	&(0.036)	&(0.034)	&(0.050)	&(0.050) \\
\hline	
Regional fixed effects	             &YES	&YES	&NO	&NO	&YES	&YES \\
R2	                                 &0.10	&0.11	&0.08	&0.09	&0.10	&0.11 \\
N	                                 &902	&1000	&914	&1016	&902	&1000 \\

\hline\hline
\end{tabular}}
\begin{tablenotes}
    \item \footnotesize{Note: Regressions estimated using survey weights. Robust standard errors are reported in parentheses. Dummy variable for ‘Primary education’ category is the reference category of the respective dummy variables set. \\
* p<0.1, ** p<0.05, *** p<0.01.}
    \item \footnotesize{Source: Own estimates based on ASFL 2019 data}
    \end{tablenotes}
\end{table}


\clearpage
\newpage
\begin{table}[!h]\centering
 \caption{Results on ESG preferences for bonds (OLS)}
 \label{table:table4}
 \resizebox{16cm}{!}{
 \begin{tabular}{lcccccc}
\hline\hline
	                                      &(M1)  	&(M2)	 &(M3)	      &(M4)	&(M5)	&(M6) \\
\hline
Preferences for enviro. issues (E)		&&	                   &-0.002	&-0.002	&-0.001	&-0.002 \\
			                            &&                      &(0.006)	&(0.006)	&(0.007)	&(0.006) \\
Preferences for social issues (S1)		&&	                    & 0.011	&	&0.011 &	\\
			                           &&                      &(0.008)	&	&(0.009) &	\\
Preferences for social issues (S2)		&&&		                       &-0.004	&	&-0.005 \\
				                         &&&                             &(0.010) &		&(0.011) \\
ESG1	                                   &0.004	&&&&&				\\
	                                     &(0.005)	&&&&&				\\
ESG2		                               &        &-0.003	 &&&&		\\	
		                                   &        &(0.006) &&&&			\\	
Objective fin. literacy	                    &0.001	&0.002	 &&		&0.002	&0.002 \\
	                                       &(0.006)	&(0.005) &&			&(0.006)	&(0.005) \\
Confidence in own fin. knowledge	       &0.013	&0.012	&&		&0.012	&0.012 \\
	                                      &(0.009)	&(0.008) &&			&(0.009)	&(0.008) \\
Risk attitude score	                    &0.044***	&0.049***	&0.047***	&0.050***	&0.045***	&0.049*** \\
	                                  &(0.012)	&(0.012)	&(0.012)	&(0.012)	&(0.012)	&(0.012) \\
Secondary education	                     &0.032*	&0.024	&0.037**	&0.031*	&0.032*	&0.025 \\
	                                     &(0.018)	&(0.016)	&(0.018)	&(0.016)	&(0.018)	&(0.016) \\
Tertiary education	                    &0.075*	&0.064	&0.085**	&0.074*	&0.076*	&0.063 \\
	                                    &(0.043)	&(0.041)	&(0.042)	&(0.041)	&(0.043)	&(0.042) \\
Individual monthly income	           &0.000***	&0.000***	&0.000***	&0.000***	&0.000***	&0.000*** \\
	                                   &(0.000)	&(0.000)	&(0.000)	&(0.000)	&(0.000)	&(0.000) \\
Gender: female	                       &-0.004	&-0.017	&-0.008	&-0.019	&-0.004	&-0.017 \\
	                                    &(0.016)	&(0.015)	&(0.016)	&(0.015)	&(0.016)	&(0.015) \\
Age	                                   &0.002***	&0.002***	&0.002***	&0.002***	&0.002***	&0.002*** \\
	                                    &(0.001)	&(0.001)	&(0.001)	&(0.001)	&(0.001)	&(0.001) \\
Constant                            	&-0.306***	&-0.242***	&-0.271***	&-0.211***	&-0.309***	&-0.241*** \\
	                                    &(0.066)	&(0.061)	&(0.055)	&(0.052)	&(0.067)	&(0.062) \\
\hline	
Regional fixed effects	                 &YES	&YES	&NO	&NO	&YES	&YES \\
R2	                                     &0.12	&0.12	&0.11	&0.11	&0.12	&0.12 \\
N	                                     &904	&998	&916	&1014	&904	&998 \\
\hline\hline
\end{tabular}}
\begin{tablenotes}
    \item \footnotesize{Note: Regressions estimated using survey weights. Robust standard errors are reported in parentheses. Dummy variable for ‘Primary education’ category is the reference category of the respective dummy variables set. \\
* p<0.1, ** p<0.05, *** p<0.01.}
    \item \footnotesize{Source: Own estimates based on ASFL 2019 data}
    \end{tablenotes}
\end{table}


\clearpage
\newpage
\begin{table}[!h]\centering
 \caption{Results on ESG preferences for shares (OLS)}
 \label{table:table5}
 \resizebox{16cm}{!}{
 \begin{tabular}{lcccccc}
\hline\hline
	                                      &(M1)  	&(M2)	 &(M3)	      &(M4)	&(M5)	&(M6) \\
\hline
Preferences for enviro. issues (E)		&&	&-0.012	&-0.011	&-0.012	&-0.012 \\
			                            && &(0.010)	&(0.010)	&(0.011)	&(0.010) \\
Preferences for social issues (S1)		&&	&0.000	&	&0.003	& \\
			                            &&  &(0.010) &	&(0.011) & \\	
Preferences for social issues (S2)		&&&		&-0.016	&	&-0.013 \\
				                        &&&    &(0.012)	&	&(0.013) \\
ESG1	                                &-0.006	&&&&& \\				
	                                    &(0.009) &&&&& \\					
ESG2		                            & &-0.012	&&&& \\			
		                                & &(0.009)	&&&& \\			
Objective fin. literacy	                &0.016*	&0.016** &&			&0.017**	&0.016** \\
	                                    &(0.009)	&(0.008)	&&		&(0.009)	&(0.008) \\
Confidence in own fin. knowledge	    &0.019	    &0.022**	&&		&0.019	    &0.022** \\
	                                    &(0.012)	&(0.011)	&&		&(0.012)	&(0.011) \\
Risk attitude score	                    &0.128***	&0.127***	&0.131***	&0.130***	&0.128***	&0.127*** \\
	                                    &(0.018)	&(0.017)	&(0.018)	&(0.017)	&(0.018)	&(0.017) \\
Secondary education	                    &0.042	    &0.027	    &0.058**	&0.047*	    &0.042	    &0.027 \\
	                                    &(0.030)	&(0.026)	&(0.029)	&(0.024)	&(0.029)	&(0.026) \\
Tertiary education	                    &0.092*	    &0.062	    &0.121**	&0.093**	&0.093*	    &0.062 \\
	                                    &(0.055)	&(0.050)	&(0.052)	&(0.047)	&(0.054)	&(0.050) \\
Individual monthly income	            &0.000***	&0.000***	&0.000***	&0.000***	&0.000***	&0.000*** \\
	                                    &(0.000)	&(0.000)	&(0.000)	&(0.000)	&(0.000)	&(0.000) \\
Gender: female	                        &0.011	    &-0.005	    &0.001	    &-0.013	    &0.011	    &-0.005 \\
	                                    &(0.020)	&(0.019)	&(0.021)	&(0.019)	&(0.020)	&(0.019) \\
Age	                                    &0.002***	&0.002***	&0.002***	&0.002***	&0.002***	&0.002*** \\
	                                    &(0.001)	&(0.001)	&(0.001)	&(0.001)	&(0.001)	&(0.001) \\
Constant	                            &-0.472***	&-0.389***	&-0.349***	&-0.269***	&-0.475***	&-0.389*** \\
	                                    &(0.111)	&(0.104)	&(0.080)	&(0.078)	&(0.110)	&(0.104) \\
\hline	
Regional fixed effects	             &YES	&YES	&NO	&NO	&YES	&YES \\
R2	                                 &0.21	&0.21	&0.20	&.19	&0.21	&0.21 \\
N	                                 &903	&1000	&915	&1016	&903	&1000 \\

\hline\hline
\end{tabular}}
\begin{tablenotes}
    \item \footnotesize{Note: Regressions estimated using survey weights. Robust standard errors are reported in parentheses. Dummy variable for ‘Primary education’ category is the reference category of the respective dummy variables set. \\
* p<0.1, ** p<0.05, *** p<0.01.}
    \item \footnotesize{Source: Own estimates based on ASFL 2019 data}
    \end{tablenotes}
\end{table}

\clearpage

\appendix
\newpage
\section*{Appendix}

\setcounter{figure}{0}
\renewcommand{\thefigure}{A.\arabic{figure}}

\setcounter{table}{0}
\renewcommand{\thetable}{A.\arabic{table}}

\begin{figure}[!h]
\caption{Share of population holding crypto-assets across Europe}
    \begin{center}
    \label{figure:figa1}
    {\includegraphics[width=15cm,keepaspectratio]{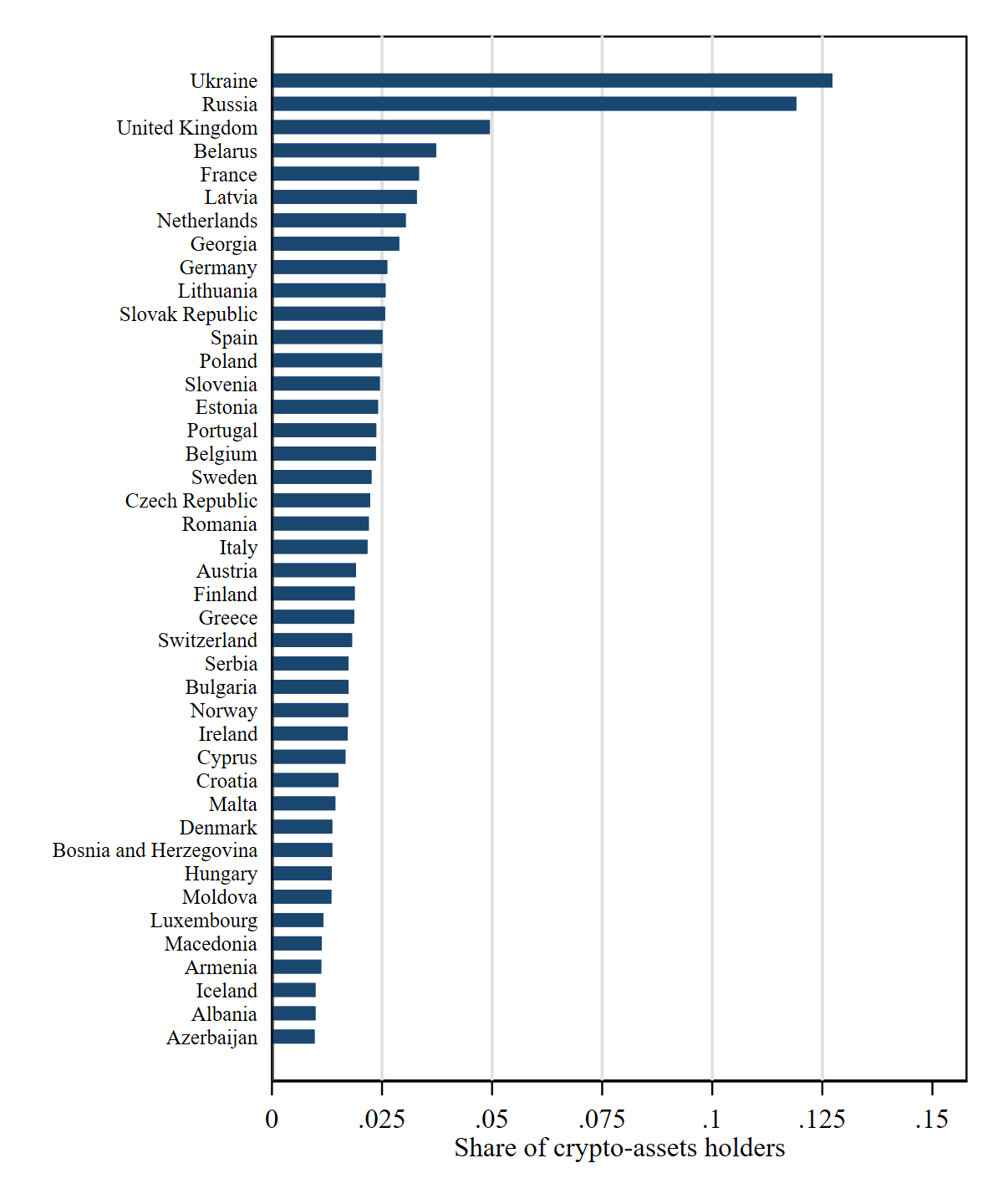}}
    \end{center}
\footnotesize{Source: Based on data from https://triple-a.io/crypto-ownership/}
\end{figure}

\clearpage
\newpage
\begin{table}[!h]
 \centering
 \caption{Description of variables used in empirical analysis}
 \label{table:tablea1}
 \resizebox{16cm}{!}{
 \begin{tabular}{lB}
 \hline\hline
 Variable                                & \multicolumn{1}{m{12cm}}{Description} \\
 \hline
 Crypto-assets ownership                & \multicolumn{1}{m{12cm}}{Dummy variable equal to 1 if an individual currently owns crypto-assets (including initial coin offerings), and 0 otherwise} \\
 Bonds ownership                           & \multicolumn{1}{m{12cm}}{Dummy variable equal to 1 if an individual currently owns bonds, and 0 otherwise} \\
 Stocks/shares ownership                 & \multicolumn{1}{m{12cm}}{Dummy variable equal to 1 if an individual currently owns stocks / shares, and 0 otherwise} \\
 Preferences for enviro. issues (E)      & \multicolumn{1}{m{12cm}}{Environmental attitudes score ranging from 1 to 5; based on the survey question: \textit{“I think it is more important for investors to choose companies that are making a profit than to choose companies that are minimising their impact on the environment”}} \\
 Preferences for social issues (S1)      & \multicolumn{1}{m{12cm}}{Social attitudes score ranging from 1 to 5; based on the survey question: \textit{“I prefer to use financial companies that have a strong ethical stance”}} \\
 Preferences for social issues (S2)      & \multicolumn{1}{m{12cm}}{Social attitudes score ranging from 1 to 5; based on the survey question: \textit{“I am honest even if it puts me at a financial disadvantage”}} \\
 ESG1 (E + S1)                          & \multicolumn{1}{m{12cm}}{Combined environmental/social score by summing E and S1 variables} \\
ESG2 (E + S2)                           & \multicolumn{1}{m{12cm}}{Combined environmental/social score by summing E and S2 variables} \\
 Objective fin. literacy                & \multicolumn{1}{m{12cm}}{Financial literacy score ranging from 0 to 7; based on correct answers to 7 financial literacy survey questions (time value of money, interest paid on loan, interest plus principal, compound interest, risk and return, definition of inflation, diversification), see \cite{oecd2018} for details} \\
 Confidence in own fin. knowledge       & \multicolumn{1}{m{12cm}}{Self-rated knowledge of financial matters ranging from 1 “very low” to 5 “very high”} \\
 Risk attitude score                           & \multicolumn{1}{m{12cm}}{Willingness to take investment risk ranging from 1 “never” to 4 “always”} \\
 Education                                 & \multicolumn{1}{m{12cm}}{Dummy variables set for the three main education categories: no or primary education, secondary education, tertiary education
} \\
Gender                      & \multicolumn{1}{m{12cm}}{Dummy variable equal to 1 if female, and 0 otherwise} \\
 Age                           & \multicolumn{1}{m{12cm}}{Age in years} \\
 Individual monthly income                            & \multicolumn{1}{m{12cm}}{Individual monthly income in euros} \\
 \hline\hline
 \end{tabular}}
    \begin{tablenotes}
    \item \footnotesize{Source: own processing based on the ASFL 2019 questionnaire}
    \end{tablenotes}
 \end{table}

\clearpage
\newpage
\begin{table}[!h]\centering
 \caption{Robustness of results on ESG preferences for crypto-assets (OLS)}
 \label{table:tablea2}
 \resizebox{16cm}{!}{
 \begin{tabular}{lcccccc}
\hline\hline
	                              &(M1)  	&(M2)	 &(M3)	      &(M4)	&(M5)	&(M6) \\
\hline
ESG1	                          &0.008*	&0.008*	 &0.008*	&&& \\		
	                              &(0.005)	&(0.005) &(0.005)	&&& \\		
ESG2				                  &&&                        &0.010**	&0.010**	&0.010** \\
				                       &&&                       &(0.005)	&(0.005)	&(0.005) \\
Objective fin. literacy	           &0.013**	&0.013**	&0.013**	&0.016***	&0.016***	&0.016***  \\
	                               &(0.006)	&(0.006)	&(0.006)	&(0.005)	&(0.005)	&(0.005) \\
Confidence in own fin. knowledge	&0.013*	&0.013*	&0.014**	&0.013*	&0.013*	&0.014** \\
	                               &(0.007)	&(0.007)	&(0.007)	&(0.007)	&(0.007)	&(0.007) \\
Risk attitude score	               &0.056***	&0.056***	&0.056***	&0.058***	&0.058***	&0.058*** \\
	                               &(0.014)	&(0.014)	&(0.014)	&(0.014)	&(0.014)	&(0.014) \\
Secondary education	               &0.004	&0.003	&0.004	&0.003	&0.002	&0.006 \\
	                               &(0.016)	&(0.015)	&(0.016)	&(0.014)	&(0.013)	&(0.015) \\
Tertiary education	              &-0.025	&-0.025	  &-0.024	    &-0.029	  &-0.030	&-0.025 \\
	                              &(0.026)	&(0.027)	&(0.026)	&(0.024)	&(0.025)	&(0.024) \\
Individual monthly income	       &-0.000	&-0.000	 &-0.000	&-0.000	&-0.000	&-0.000 \\
	                               &(0.000)	&(0.000)	&(0.000)	&(0.000)	&(0.000)	&(0.000) \\
Individual monthly income squared	&	&-0.000	&&		&-0.000	 & \\
		                            &  &(0.000)	&&		&(0.000) & \\	
Gender: female	                    &-0.008	&-0.008	&-0.008	&0.004	&0.004	&0.004 \\
	                                &(0.012)	&(0.012)	&(0.012)	&(0.013)	&(0.013)	&(0.013) \\
Age	                                &-0.001**	&-0.001**	&-0.001	&-0.001**	&-0.001**	&-0.003 \\
	                                &(0.000)	&(0.000)	&(0.002)	&(0.000)	&(0.000)	&(0.002) \\
Age squared			                &&                    &0.000	&&		&0.000 \\
			                        &&                    &(0.000)	&&		&(0.000) \\
Constant                        &-0.155***	&-0.155***	&-0.148***	&-0.175***	&-0.177***	&-0.145*** \\
	                            &(0.049)	&(0.051)	&(0.056)	&(0.049)	&(0.049)	&(0.053) \\
\hline
Regional fixed effects	         &YES	&YES	&YES	&YES	&YES	&YES \\
R2	                              &0.10	&0.10	&0.10	&0.11	&0.11	&0.11 \\
N	                             &902	&902	&902	&1000	&1000	&1000 \\
\hline\hline
\end{tabular}}
\begin{tablenotes}
    \item \footnotesize{Note: Regressions estimated using survey weights. Robust standard errors are reported in parentheses. Dummy variable for ‘Primary education’ category is the reference category of the respective dummy variables set. \\
* p<0.1, ** p<0.05, *** p<0.01.}
    \item \footnotesize{Source: Own estimates based on ASFL 2019 data}
    \end{tablenotes}
\end{table}

\clearpage
\newpage
\begin{table}[!h]\centering
 \caption{Robustness of results on ESG preferences for crypto-assets (OLS, probit, and rare-events logit models)}
 \label{table:tablea3}
 \resizebox{16cm}{!}{
 \begin{tabular}{lFFFFFF}
\hline\hline
	                              &(M1)  	&(M2)	 &(M3)	      &(M4)	&(M5)	&(M6) \\
	                               &OLS  	&probit	 &rare-events logit	&OLS  	&probit	 &rare-events logit \\
\hline
ESG1	                           &0.008*	&0.002	&0.006		&&& \\	
	                               &(0.005)	&(0.002)	&(0.006) &&& \\			
ESG2			                   &&&	                    &0.009**	&0.003*	&0.009* \\
				                   &&&                      &(0.004)	&(0.002) &(0.005) \\
Objective fin. literacy	          &0.014**	&0.004**	&0.014*	&0.017***	&0.005***	&0.018*** \\
	                              &(0.006)	&(0.002)	&(0.007)	&(0.006)	&(0.002)	&(0.007) \\
Confidence in own fin. knowledge	&0.012*	&0.008**	&0.018*	&0.012*	&0.006*	&0.015* \\
	                                &(0.007)	&(0.004)	&(0.009)	&(0.007)	&(0.003)	&(0.008) \\
Risk attitude score	                &0.058***	&0.016***	&0.040***	&0.061***	&0.013***	&0.042*** \\
	                                &(0.014)	&(0.005)	&(0.009)	&(0.014)	&(0.005)	&(0.009) \\
Secondary education	                &0.003	    &0.003	    &-0.000	    &0.002	    &0.006	    &0.012 \\
                                  	&(0.016)	&(0.014)	&(0.046)	&(0.014)	&(0.012)	&(0.053) \\
Tertiary education	               &-0.023	&-0.008	&-0.029	&-0.027	&-0.003	&-0.017 \\
	                               &(0.026)	&(0.016)	&(0.051)	&(0.024)	&(0.013)	&(0.055) \\
Individual monthly income	       &-0.000	&-0.000	&-0.000	&-0.000	&-0.000	&-0.000 \\
                                 	&(0.000)	&(0.000)	&(0.000)	&(0.000)	&(0.000)	&(0.000) \\
Gender: female	                   &-0.008	&-0.001	&-0.005	&0.004	&0.004	&0.009 \\
                                 	&(0.012)	&(0.005)	&(0.015)	&(0.013)	&(0.005)	&(0.014) \\
Age	                               &-0.001**	&-0.001***	&-0.001***	&-0.001**	&-0.000***	&-0.002*** \\
	                               &(0.000)	&(0.000)	&(0.001)	&(0.000)	&(0.000)	&(0.001) \\
\hline
Regional fixed effects	&YES	&YES	&YES	&YES	&YES	&YES \\
R2	                   &0.10	&&		&0.11 && \\		
Pseudo R2	           &	&0.27 &&    &0.30	& \\
N	                   &902	&902	&902	&1000	&1000	&1000 \\
\hline\hline
\end{tabular}}
\begin{tablenotes}
    \item \footnotesize{Note: For probit and rare-events logit models we report marginal effects (calculated at the means of explanatory variables). Rare-events logit models are estimated using ‘relogit’ Stata estimation command \citep{tomz2021}. Regressions estimated using survey weights. Robust standard errors are reported in parentheses. Dummy variable for ‘Primary education’ category is the reference category of the respective dummy variables set. \\
* p<0.1, ** p<0.05, *** p<0.01.}
    \item \footnotesize{Source: Own estimates based on ASFL 2019 data}
    \end{tablenotes}
\end{table}

\end{document}